\documentclass[aps]{PoS}

\def\eq#1{(\ref{#1})}

\def\s0#1#2{\mbox{\small{$ \frac{#1}{#2} $}}}
\def\0#1#2{\frac{#1}{#2}}

\def\beq{\begin{equation}}
\def\eeq{\end{equation}}

\def\bea{\arraycolsep .1em \begin{eqnarray}}
\def\eea{\end{eqnarray}}
\def\tr{{\rm Tr}}

\title{Fixed Points of Quantum Gravity and the Renormalisation Group}

\ShortTitle{Fixed Points of Quantum Gravity\hskip8.6cm 
{\rm Daniel Litim}}

\author{Daniel F. Litim\\
  Department of Physics and Astronomy, University of Sussex, Brighton, BN1
  9QH, U.K.\\
  E-mail: \email{d.litim@sussex.ac.uk}
}

\abstract{We review the asymptotic safety scenario for quantum gravity and the
  role and implications of an underlying ultraviolet fixed point.  We discuss
  renormalisation group techniques employed in the fixed point search, analyse
  the main picture at the example of the Einstein-Hilbert theory, and provide
  an overview of the key results in four and higher dimensions. We also
  compare findings with recent lattice simulations and evaluate
  phenomenological implications for collider experiments.}

\FullConference{From Quantum to Emergent Gravity: Theory and Phenomenology\\
                June 11-15 2007\\
                Trieste, Italy}

\begin{document}

\section{Introduction}
It is commonly believed that an understanding of the dynamics of gravity and
the structure of space-time at shortest distances requires an explicit quantum
theory for gravity. The well-known fact that the perturbative quantisation
program for gravity in four dimensions faces problems has raised the suspicion
that a consistent formulation of the theory may require a radical deviation
from the concepts of local quantum field theory, $e.g.$ string theory.
It remains an interesting and open challenge to prove, or falsify, that a
consistent quantum theory of gravity cannot be accommodated for within the
otherwise very successful framework of local quantum field theories.

Some time ago Steven Weinberg added a new perspective to this problem by
pointing out that a quantum theory of gravity in terms of the metric field may
very well exist, and be renormalisable on a non-perturbative level, despite
it's notorious perturbative non-renormalisability \cite{Weinberg}. This
scenario, since then known as ``asymptotic safety'', necessitates an
interacting ultraviolet fixed point for gravity under the renormalisation
group (RG)
\cite{Weinberg,Litim:2006dx,Niedermaier:2006ns,NiedermaierReuter,Percacci:2007sz}. If
so, the high energy behaviour of gravity is governed by near-conformal scaling
in the vicinity of the fixed point in a way which circumnavigates the virulent
ultraviolet (UV) divergences encountered within standard perturbation theory.
Indications in favour of an ultraviolet fixed point are based on
renormalisation group studies in four and higher dimensions
\cite{Reuter:1996cp,Souma:1999at,Lauscher:2001ya,Lauscher:2002mb,Reuter:2001ag,Percacci:2002ie,Litim:2003vp,Bonanno:2004sy,Fischer:2006fz,Fischer:2006at,DL06,Codello:2007bd,Machado:2007ea},
dimensional reduction techniques \cite{Forgacs:2002hz,Niedermaier:2002eq},
renormalisation group studies in lower dimensions
\cite{Weinberg,Gastmans:1977ad,Christensen:1978sc,Kawai:1992fz,Aida:1996zn},
four-dimensional perturbation theory in higher derivative gravity
\cite{Codello:2006in}, large-$N$ expansions in the matter fields
\cite{Percacci:2005wu}, and lattice simulations
\cite{Hamber:1999nu,Hamber:2005vc,Ambjorn:2004qm}.

In this contribution, we review the key elements of the asymptotic safety
scenario (Sec.~\ref{AS}) and introduce renormalisation group techniques
(Sec.~\ref{RG}) which are at the root of fixed point searches in quantum
gravity. The fixed point structure in four (Sec.~\ref{FP}) and
higher dimensions (Sec.~\ref{ED}), and the phase diagram of gravity
(Sec.~\ref{PD}) are discussed and evaluated in the light of the underlying
approximations. Results are compared with recent lattice simulations
(Sec.~\ref{Lattice}), and phenomenological implications are indicated
(Sec.~\ref{pheno}). We close with some conclusions (Sec.~\ref{conclusions}).

\section{Asymptotic Safety}
\label{AS}
We summarise the basic set of ideas and assumptions of asymptotic safety as
first laid out in \cite{Weinberg} (see
\cite{Litim:2006dx,Niedermaier:2006ns,NiedermaierReuter,Percacci:2007sz} for
reviews).  The aim of the asymptotic safety scenario for gravity is to provide
for a path-integral based framework in which the metric field is the carrier
of the fundamental degrees of freedom, both in the classical and in the
quantum regimes of the theory. This is similar in spirit to effective field
theory approaches to quantum gravity \cite{Donoghue:1993eb}. There, a
systematic study of quantum effects is possible without an explicit knowledge
of the ultraviolet completion as long as the relevant energy scales are much
lower than the ultraviolet cutoff $\Lambda$ of the effective theory, with
$\Lambda$ of the order of the Planck scale (see \cite{Burgess:2003jk} for a
recent review).

The asymptotic safety scenario goes one step further and assumes that the
cutoff $\Lambda$ can in fact be removed, $\Lambda\to\infty$, and that the
high-energy behaviour of gravity, in this limit, is characterised by an
interacting fixed point.  It is expected that the relevant field
configurations dominating the gravitational path integral at high energies are
predominantly ``anti-screening'' to allow for this limit to become
feasible. If so, it is conceivable that a non-trivial high-energy fixed point
of gravity may exist and should be visible within $e.g.$ renormalisation group
or lattice implementations of the theory, analogous to the well-known
perturbative high-energy fixed point of QCD.  Then the high-energy behaviour
of the relevant gravitational couplings is ``asymptotically safe'' and
connected with the low-energy behaviour by finite renormalisation group flows.
The existence of a fixed point together with finite renormalisation group
trajectories provides for a definition of the theory at arbitrary energy
scales.

The fixed point implies that the high-energy behaviour of gravity is
characterised by universal scaling laws, dictated by the residual high-energy
interactions. No a priori assumptions are made about which invariants are the
relevant operators at the fixed point. In fact, although the low-energy
physics is dominated by the Einstein-Hilbert action, it is expected that (a
finite number of) further invariants will become relevant, in the
renormalisation group sense, at the ultraviolet fixed point.\footnote{For
  {infrared} fixed points, universality considerations often simplify the task
  of identifying the set of relevant, marginal and irrelevant operators. This
  is not applicable for interacting {ultraviolet} fixed points.} Then, in
order to connect the ultraviolet with the infrared physics along some
renormalisation group trajectory, a finite number of initial parameters have
to be fixed, ideally taken from experiment. In this light, classical general
relativity would emerge as a ``low-energy phenomenon'' of a fundamental
quantum field theory in the metric field.

We illustrate this scenario with a discussion of the renormalisation group
equation for the gravitational coupling $G$, following \cite{Litim:2006dx}
(see also \cite{Niedermaier:2006ns,NiedermaierReuter}). Its canonical
dimension is $[G]=2-d$ in $d$ dimensions and hence negative for $d>2$.  It is
commonly believed that a negative mass dimension for the relevant coupling is
responsible for the perturbative non-renormalisability of the theory.  We
introduce the renormalised coupling as $G(\mu)=Z^{-1}_G(\mu)\, G$, and the
dimensionless coupling as $g(\mu)=\mu^{d-2}\,G(\mu)$; the momentum scale $\mu$
denotes the renormalisation scale. The graviton wave function renormalisation
factor $Z_G(\mu)$ is normalised as $Z_G(\mu_0)=1$ at 
$\mu=\mu_0$ with $G(\mu_0)$ given by Newton's coupling constant $G_N=6.67428
\cdot 10^{-11}\s0{m^3}{{\rm kg}\,s^2}$.  The graviton anomalous dimension
$\eta$ related to $Z_G(\mu)$ is given by $\eta=-\mu\s0{{\rm d}}{{\rm d}\mu}
\ln Z_G$. Then the Callan-Symanzik equation for $g(\mu)$ reads
\begin{equation}\label{dg}
\beta_g\equiv\mu\frac{{\rm d}g(\mu)}{{\rm d}\mu}
=(d-2+\eta)g(\mu)\,.
\end{equation}
Here we have assumed a fundamental action for gravity which is local in the
metric field. In general, the graviton anomalous dimension $\eta(g,\cdots)$ is
a function of all couplings of the theory including matter fields. The RG
equation \eq{dg} displays two qualitatively different types of fixed
points. The non-interacting (gaussian) fixed point corresponds to $g_*=0$
which also entails $\eta=0$. In its vicinity with $g(\mu_0)\ll 1$, we have
canonical scaling since $\beta_g=(d-2)g$, and
\begin{equation}
\label{Gauss}
G(\mu) = G(\mu_0)
\end{equation}
for all $\mu<\mu_0$. Consequently, the gaussian regime corresponds to the
domain of classical general relativity. In turn, \eq{dg} can display an
interacting fixed point $g_*\neq 0$ in $d>2$ if the anomalous dimension takes
the value $\eta(g_*,\cdots)=2-d$; the dots denoting further gravitational and
matter couplings.  Hence, the anomalous dimension precisely counter-balances
the canonical dimension of Newton's coupling $G$.  This structure is at the
root for the non-perturbative renormalisability of quantum gravity within a
fixed point scenario.\footnote{Integer values for anomalous dimensions are
  well-known from other gauge theories at criticality and away from their
  canonical dimension. In the $d$-dimensional $U(1)$+Higgs theory, the abelian
  charge $e^2$ has mass dimension $[e^2]=4-d$, with $\beta_{e^2}=(d-4+\eta)\,
  e^2$. In three dimensions, a non-perturbative infrared fixed point at
  $e^2_*\neq 0$ leads to $\eta_*=1$ \cite{Bergerhoff:1995zq}. The fixed point
  belongs to the universality class of conventional superconductors with the
  charged scalar field describing the Cooper pair. The integer value $\eta_*
  =1$ implies that the magnetic field penetration depth and the Cooper pair
  correlation length scale with the same universal exponent at the phase
  transition \cite{Bergerhoff:1995zq,Herbut:1996ut}.  In Yang-Mills theories
  above four dimensions, ultraviolet fixed points with $\eta=4-d$ and
  implications thereof have been discussed in
  \cite{Kazakov:2002jd,Gies:2003ic,Morris:2004mg}.}  Consequently, at an
interacting fixed point where $g_*\neq0$, the anomalous dimension implies the
scaling
\begin{equation}\label{G-asym}
G(\mu)= \frac{g_*}{\mu^{d-2}}
\end{equation}
for the dimensionful gravitational coupling. In the case of an ultraviolet
fixed point $g_*\neq 0$ for large $\mu$, the dimensionful coupling $G$ becomes
arbitrarily small in its vicinity. This is in marked contrast to
\eq{Gauss}. Hence, \eq{G-asym} indicates that gravity weakens at the onset of
fixed point scaling.  Nevertheless, at the fixed point the theory remains
non-trivially coupled because of $g_*\neq 0$. The weakness of the coupling in
\eq{G-asym} is a dimensional effect, and should be contrasted with $e.g.$
asymptotic freedom of QCD in four dimensions where the dimensionless
non-abelian gauge coupling becomes weak because of a non-interacting
ultraviolet fixed point. In turn, if \eq{G-asym} corresponds to a non-trivial
infrared fixed point for $\mu\to 0$, the dimensionful coupling $G(\mu)$ grows
large. A strong coupling behaviour of this type would imply interesting long
distance modifications of gravity. 

As a final comment, we point out that asymptotically safe gravity is expected
to become, in an essential way, two-dimensional at high energies.
Heuristically, this can be seen from the dressed graviton propagator whose
scalar part, neglecting the tensorial structure, scales as ${\cal G}(p^2)\sim
p^{-2(1-\eta/2)}$ in momentum space. Here we have evaluated the anomalous
dimension at $\mu^2\approx p^2$. Then, for small $\eta$, we have the standard
perturbative behaviour $\sim p^{-2}$. In turn, for large anomalous dimension
$\eta\to 2-d$ in the vicinity of a fixed point the propagator is additionally
suppressed $\sim (p^2)^{-d/2}$ possibly modulo logarithmic corrections. After
Fourier transform to position space, this corresponds to a logarithmic
behaviour for the propagator ${\cal G}(x,y)\sim \ln(|x-y|\mu)$, characteristic
for bosonic fields in two-dimensional systems.

\section{Renormalisation Group}\label{Implications}
\label{RG}
Whether or not a non-trivial fixed point is realised in quantum gravity can be
assessed once explicit renormalisation group equations for the scale-dependent
gravitational couplings are available. To that end, we recall the set-up of
Wilson's (functional) renormalisation group (see
\cite{Litim:1998yn,Litim:1998nf,Bagnuls:2000ae,BTW,JP,Pawlowski:2005xe,Gies:2006wv}
for reviews), which is used below for the case of quantum gravity. Wilsonian
flows are based on the notion of a cutoff effective action $\Gamma_k$, where
the propagation of fields $\phi$ with momenta smaller than $k$ is suppressed.
A Wilsonian cutoff is realised by adding $\Delta
S_k=\s012\int\varphi(-q)\,R_k(q)\,\varphi(q)$ within the Schwinger functional
\begin{equation}\label{Z}
\ln\, Z_k[J]=\ln \int [D\varphi]_{\rm ren.}\exp\left(-S[\varphi]
-\Delta S_k[\varphi]
+\int J\cdot \varphi
\right)\,
\end{equation}
and the requirement that $R_k$ obeys (i) $R_k(q)\to 0$ for $k^2/q^2\to 0$,
(ii) $R_k(q)> 0$ for $q^2/k^2\to 0$, and (iii) $R_k(q)\to\infty$ for
$k\to\Lambda$ (for examples and plots of $R_k$, see \cite{Litim:2000ci}). Note
that the Wilsonian momentum scale $k$ takes the role of the renormalisation
group scale $\mu$ introduced in the previous section. Under infinitesimal
changes $k\to k-\Delta k$, the Schwinger functional obeys $\partial_t\ln
Z_k=-\langle\partial_t\Delta S_k\rangle_J$; $t=\ln k$. We also introduce its
Legendre transform, the scale-dependent effective action
$\Gamma_k[\phi]=\sup_J\left(\int J\cdot \phi -\ln Z_k[J]\right)-\s012\int \phi
R_k\phi$, $\phi=\langle \varphi\rangle_J$. It obeys an exact functional
differential equation introduced by Wetterich \cite{Wetterich:1992yh}
\beq\label{ERG} 
\partial_t \Gamma_k=
\frac{1}{2} \tr \left({\Gamma_k^{(2)}+R_k}\right)^{-1}\partial_t R_k\,,
\eeq 
which relates the change in $\Gamma_k$ with a one-loop type integral over the
full field-dependent cutoff propagator. Here, the trace $\tr$ denotes an
integration over all momenta and summation over all fields, and
$\Gamma_k^{(2)}[\phi](p,q)\equiv \delta^2\Gamma_k/\delta\phi(p)\delta\phi(q)$.
A number of comments are in order:
\begin{itemize}
\item[$\bullet$] {\bf Finiteness and interpolation property.} By construction,
  the flow equation \eq{ERG} is well-defined and finite, and interpolates
  between an initial condition $\Gamma_\Lambda$ for $k\to \Lambda$ and the
  full effective action $\Gamma\equiv\Gamma_{k=0}$. This is illustrated in
  Fig.~\ref{Vergleich}. The endpoint is independent of the regularisation,
  whereas the trajectories $k\to \Gamma_k$ depend on it.
\item[$\bullet$] {\bf Locality.} The integrand of \eq{ERG} is peaked for field
  configurations with momentum squared $q^2\approx k^2$, and suppressed for
  large momenta [due to condition (i) on $R_k$] and for small momenta [due to
    condition (ii)]. Therefore, the flow equation is essentially local in
  momentum and field space \cite{Litim:2000ci,Litim:2005us}.
\item[$\bullet$] {\bf Approximations.} Systematic approximations for
  $\Gamma_k$ and $\partial_t\Gamma_k$ are required to integrate
  \eq{ERG}. These include (a) perturbation theory, (b) expansions in powers of
  the fields (vertex functions), (c) expansion in powers of derivative
  operators (derivative expansion), and (d) combinations thereof. The
  iterative structure of perturbation theory is fully reproduced to all
  orders, independently of $R_k$ \cite{Litim:2001ky,Litim:2002xm}. The
  expansions (b) - (d) are genuinely non-perturbative and lead, via \eq{ERG}, to
  coupled flow equations for the coefficient functions. Convergence is then
  checked by extending the approximation to higher order.
\item[$\bullet$] {\bf Stability.} The stability and convergence of
  approximations is, additionally, controlled by $R_k$
  \cite{Litim:2000ci,Litim:2001up}. Here, powerful optimisation techniques are
  available to maximise the physics content and the reliability through
  well-adapted choices of $R_k$
  \cite{Litim:2000ci,Litim:2001up,Litim:2005us,Pawlowski:2003hq,Pawlowski:2005xe}. These
  ideas have been explicitly tested in $e.g.$~scalar \cite{Litim:2002cf} and
  gauge theories \cite{Pawlowski:2003hq}.
\item[$\bullet$] {\bf Symmetries.} Global or local (gauge/diffeomorphism)
  symmetries of the underlying theory can be expressed as Ward-Takahashi
  identities for $n$-point functions of $\Gamma$. Ward-Takahashi identities
  are maintained for all $k$ if the insertion $\Delta S_k$ is compatible with
  the symmetry. In general, this is not the case for non-linear symmetries
  such as in non-Abelian gauge theories or gravity. Then the requirements of
  gauge symmtry for $\Gamma$ are preserved by either (a) imposing modified
  Ward identities which ensure that standard Ward identities are obeyed in the
  the physical limit when $k\to 0$, or by (b) introducing background fields
  into the regulator $R_k$ and taking advantage of the background field
  method, or by (c) using gauge-covariant variables rather than the gauge
  fields or the metric field \cite{Branchina:2003ek}. For a discussion of
  benefits and shortcomings of these options see
  \cite{Litim:1998nf,Pawlowski:2005xe}. For gravity, most implementations
  presently employ option (b) together with optimisation techniques to control
  the symmetry \cite{Fischer:2006fz,Fischer:2006at}.

\item[$\bullet$] {\bf Integral representation.} The physical theory described
  by $\Gamma$ can be defined without explicit reference to an underlying
  path integral representation, using only the (finite) initial condition
  $\Gamma_\Lambda$, and the (finite) flow equation
  \eq{ERG}
\begin{figure}[t]
\begin{center}
  \unitlength0.001\hsize
\begin{picture}(1000,260)
\put(50,250){{a)}}
\put(500,250){{b)}}
\put(220,220){$S$}
\put(680,220){$\Gamma_*$}
\put(210,130){$k\partial_k\Gamma_{k}$}
\put(690,130){$k\partial_k\Gamma_{k}$}
\put(180,-20){$\Gamma_{0}\approx\Gamma$}
\put(660,-20){$\Gamma_{0}\approx\Gamma_{\rm EH}$}
{}\hskip.05\hsize
\includegraphics[width=.42\hsize]{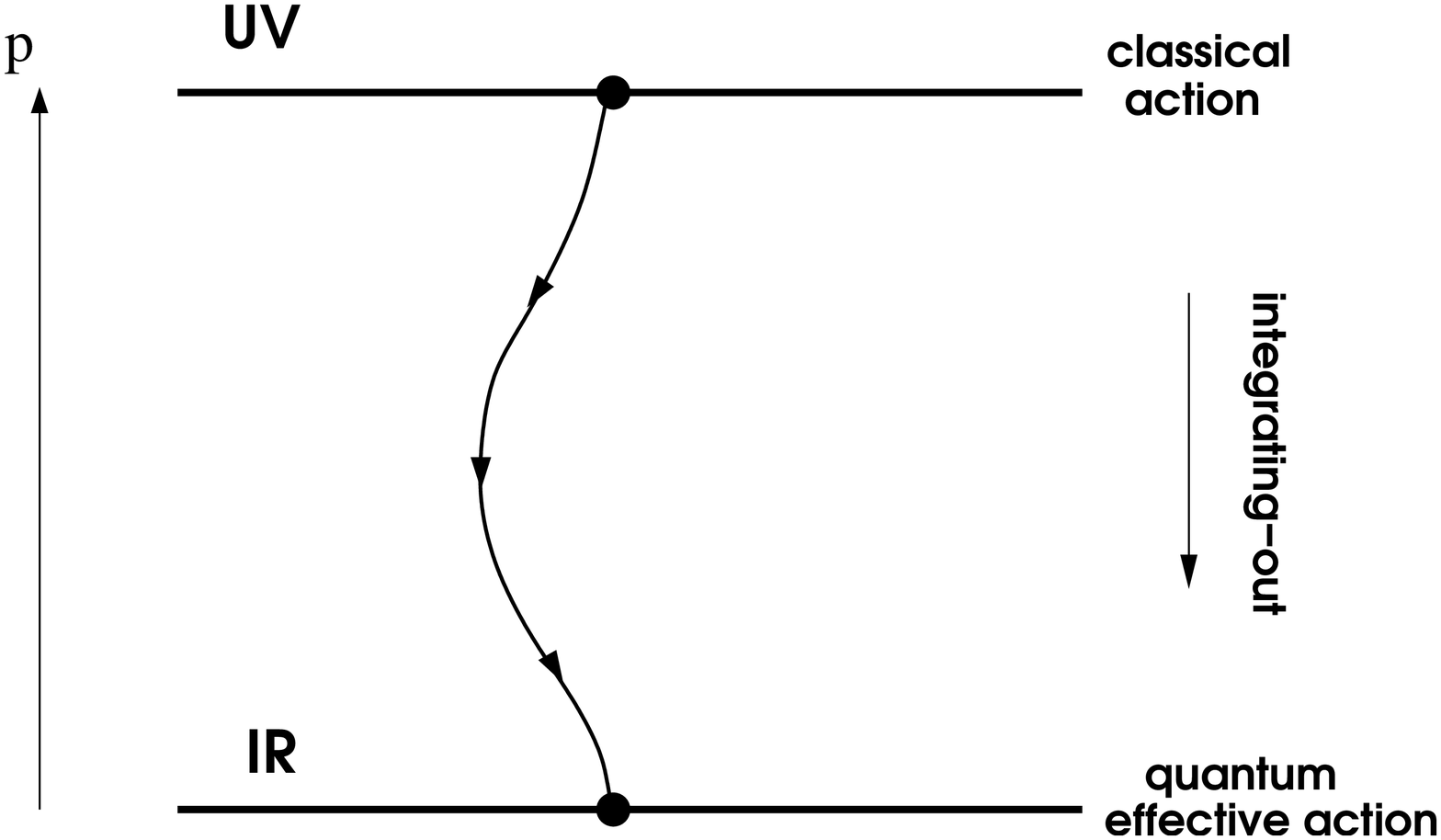}
\hskip.05\hsize
\includegraphics[width=.42\hsize]{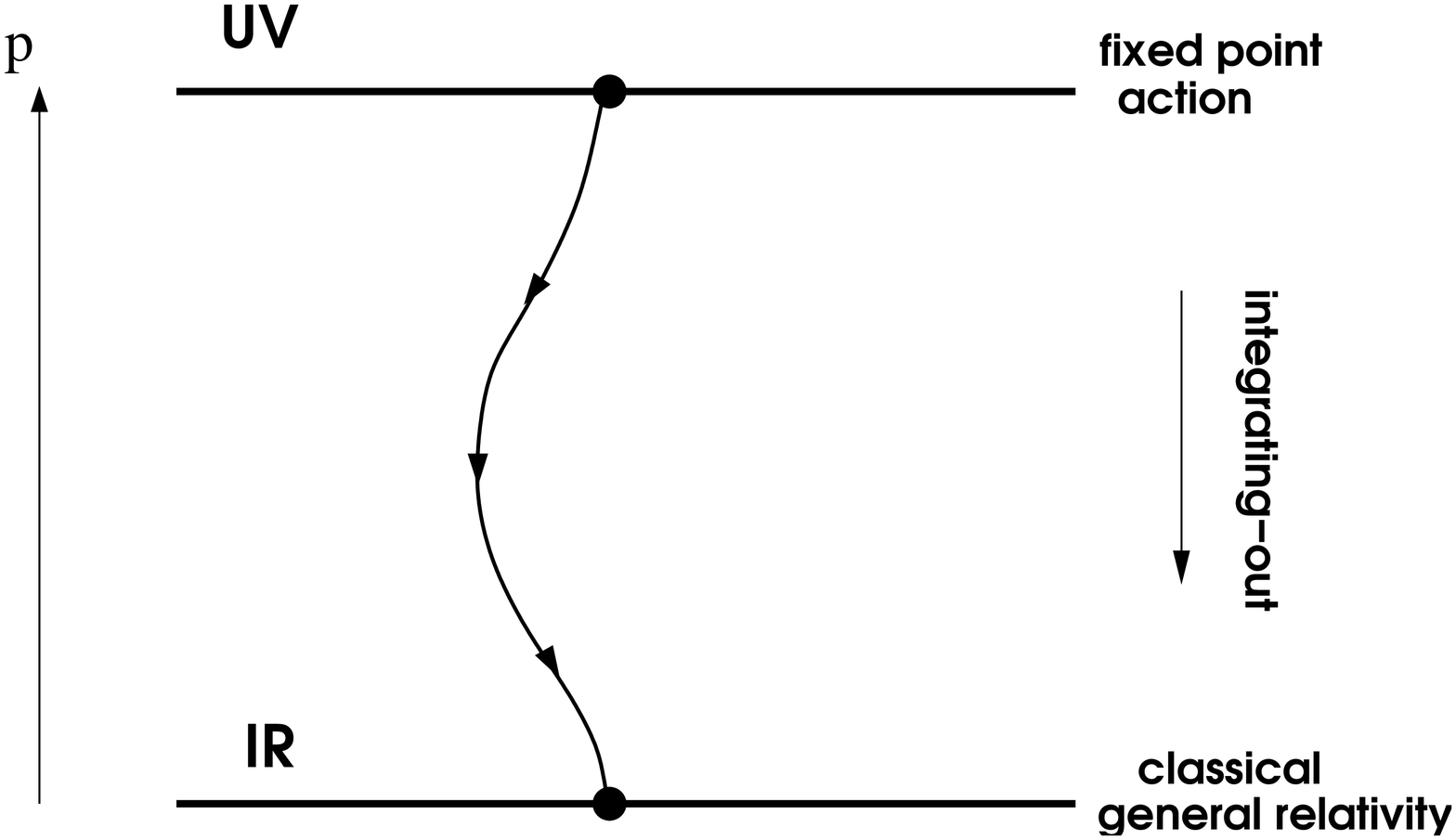}
\end{picture}
\vskip.5cm
\caption{\label{Vergleich} Wilsonian flows for scale-dependent effective
  actions $\Gamma_k$ in the space of all action functionals (schematically);
  arrows point towards smaller momentum scales and lower energies $k\to 0$.
  {a)} Flow connecting a fundamental classical action $S$ at high energies
  in the ultraviolet with the full quantum effective action $\Gamma$ at low
  energies in the infrared (``top-down''). {b)} Flow connecting the
  Einstein-Hilbert action at low energies with a fundamental fixed point
  action $\Gamma_*$ at high energies (``bottom-up'').}
\end{center}
\end{figure}
\begin{equation}\label{integral}
\Gamma=\Gamma_\Lambda+
\int_\Lambda^0
\s0{dk}{k}\s012\,
\tr \left({\Gamma_k^{(2)}+R_k}\right)^{-1}\partial_t R_k\,.
\end{equation}
This provides an implicit regularisation of the path integral
underlying \eq{Z}. It should be compared with the standard representation
for $\Gamma$ via a functional integro-differential equation
\begin{equation}\label{DSE}
e^{-\Gamma}
=\int [D\varphi]_{\rm ren.}\exp
\left(-S[\phi+\varphi]
+\int \0{\delta\Gamma[\phi]}{\delta\phi}\cdot\varphi\right)
\end{equation}
which is at the basis of $e.g.$ the hierarchy of Dyson-Schwinger equations.
\item[$\bullet$] {\bf Renormalisability.} In renormalisable theories, the
  cutoff $\Lambda$ in \eq{integral} can be removed, $\Lambda\to\infty$, and
  $\Gamma_\Lambda\to \Gamma_*$ remains well-defined for arbitrarily short
  distances. In perturbatively renormalisable theories, $\Gamma_*$ is given by
  the classical action $S$, such as in QCD.  In this case, illustrated in
  Fig.~\ref{Vergleich}a), the high energy behaviour of the theory is simple,
  given mainly by the classical action, and the challenge consists in deriving
  the physics of the strongly coupled low energy limit.  In perturbatively
  non-renormalisable theories such as quantum gravity, proving the existence
  (or non-existence) of a short distance limit $\Gamma_*$ is more difficult.
  For gravity, illustrated in Fig.~\ref{Vergleich}b), experiments indicate
  that the low energy theory is simple, mainly given by the Einstein Hilbert
  theory. The challenge consists in identifying a possible high energy fixed
  point action $\Gamma_*$, which upon integration matches with the known
  physics at low energies.  In principle, any $\Gamma_*$ with the above
  properties qualifies as fundamental action for quantum gravity. In
  non-renormalisable theories the cutoff $\Lambda$ cannot be removed. Still,
  the flow equation allows to access the physics at all scales $k<\Lambda$
  analogous to standard reasoning within effective field theory
  \cite{Burgess:2003jk}.
\item[$\bullet$] {\bf Link with Callan-Symanzik equation.} The well-known
  Callan-Symanzik equation describes a flow $k\s0d{dk}$ driven by a mass
  insertion $\sim k^2\phi^2$. In \eq{ERG}, this corresponds to the choice
  $R_k(q^2)=k^2$, which does not fulfill condition (i). Consequently, the
  corresponding flow is no longer local in momentum space, and requires an
  additional UV regularisation. This highlights a crucial difference between
  the Callan-Symanzik equation and functional flows \eq{ERG}. In this light,
  the flow equation \eq{ERG} could be interpreted as a functional
  Callan-Symanzik equation with {momentum-dependent} mass term insertion
  \cite{Symanzik:1970rt}.
\end{itemize}

Now we are in a position to implement these ideas for quantum gravity
\cite{Reuter:1996cp}. A Wilsonian effective action for gravity $\Gamma_k$
should contain the Ricci scalar $R(g_{\mu\nu})$ with a running gravitational
coupling $G_k$, a running cosmological constant $\Lambda_k$ (with canonical
mass dimension $[\Lambda_k]=2$), possibly higher order interactions in the
metric field such as powers, derivatives, or functions of $e.g.$ the Ricci
scalar, the Ricci tensor, the Riemann tensor, and, possibly, non-local
operators in the metric field. The effective action should also contain a
standard gauge-fixing term $S_{\rm gf}$, a ghost term $S_{\rm gh}$ and matter
interactions $S_{\rm matter}$. Altogether,
\begin{equation}\label{EHk}
\Gamma_k=
\int d^dx \sqrt{\det g_{\mu\nu}}\left[
\0{1}{16\pi G_k}\left(-R+2\Lambda_k \right)
+\cdots+S_{\rm gf}+S_{\rm gh}+ S_{\rm matter}\right]\,,
\end{equation}
and explicit flow equations for the coefficient functions such as $G_k$,
$\Lambda_k$ or vertex functions, are obtained by appropriate projections after
inserting \eq{EHk} into \eq{ERG}.  All couplings in \eq{EHk} become running
couplings as functions of the momentum scale $k$.  For $k$ much smaller than
the $d$-dimensional Planck scale $M_*$, the gravitational sector is
well approximated by the Einstein-Hilbert action with $G_k\approx G_{k=0}$,
and similarily for the gravity-matter couplings.  At $k\approx M_*$ and above,
the RG running of gravitational couplings becomes important. This
is the topic of the following sections.

A few technical comments are in order: To ensure gauge symmetry within this
set-up, we take advantage of the background field formalism and add a
non-propagating background field $\bar g_{\mu\nu}$
\cite{Reuter:1996cp,Litim:1998nf,Reuter:1993kw,Litim:1998qi,Freire:2000bq,Litim:2002hj,Pawlowski:2001df}. This
way, the extended effective action $\Gamma_k[g_{\mu\nu},\bar g_{\mu\nu}]$
becomes gauge-invariant under the combined symmetry transformations of the
physical and the background field. A second benefit of this is that the
background field can be used to construct a covariant Laplacean $-\bar D^2$,
or similar, to define a mode cutoff at momentum scale $k^2=-\bar D^2$. This
implies that the mode cutoff $R_k$ will depend on the background fields. The
background field is then eliminated from the final equations by identifying it
with the physical mean field. This procedure, which dynamically readjusts the
background field, implements the requirements of ``background independence''
for quantum gravity. For a detailed evaluation of Wilsonian background field
flows, see \cite{Litim:2002hj}. Finally, we note that the operator traces
$\tr$ in \eq{ERG} are evaluated using heat kernel techniques. Here,
well-adapted choices for $R_k$ \cite{Litim:2000ci,Litim:2001up} lead to
substantial algebraic simplifications, and open a door for systematic fixed
point searches, which we discuss next.

\section{Fixed Points}
\label{FP}
In this section, we discuss the main picture in a simple approximation which
captures the salient features of an asymptotic safety scenario for gravity,
and give an overview of extensions. We consider the Einstein-Hilbert theory
with a cosmological constant term and employ a momentum cutoff $R_k$ with the
tensorial structure of \cite{Lauscher:2001ya} and variants thereof, an
optimised scalar cutoff $R_k(q^2)\sim(k^2-q^2)\theta(k^2-q^2)$
\cite{Litim:2000ci,Litim:2001up}, and a harmonic background field gauge with
parameter $\alpha$ in a specific limit introduced in \cite{Litim:2003vp}.  The
ghost wave function renormalisation is set to $Z_{C,k}=1$, and the effective
action is given by \eq{EHk} with $S_{\rm matter}=0$.  In the domain of
classical scaling $G_k$ and $\Lambda_k$ are approximately constant, and
\eq{EHk} reduces to the conventional Einstein-Hilbert action in $d$ euclidean
dimensions. The dimensionless renormalised gravitational and cosmological
constants are
\begin{equation}\label{glk}
\begin{array}{l}
g=k^{d-2}\, G_k\, \equiv k^{d-2}\, Z^{-1}_{G}(k)\ \bar G\ ,
\quad\quad
\lambda=\,k^{-2}\, \Lambda_k\ 
\end{array}
\end{equation}
where it is understood that $g$ and $\lambda$ depend on $k$.  Then the coupled
system of $\beta$-functions is
\begin{eqnarray}
\partial_t\lambda\equiv\beta_{\lambda}(\lambda,g)&=&
-2\lambda +\frac{g}{2}\,d\,(d+2)\,(d-5)
-d(d+2)g\, \frac{(d-1)g
+\frac{1}{d-2}(1-4\frac{d-1}{d}\lambda)}{2g-\frac{1}{d-2}(1-2\lambda)^2}
\label{beta-l-inf}
\\
\label{beta-g-inf}
\partial_t g\equiv\beta_{ g}(\lambda,g)&=&
(d-2)g+\frac{2(d-2)(d+2)g^2}{2(d-2)g-(1-2\lambda)^2}\,.
\end{eqnarray}
We have rescaled $g\to g/c_d$ with $c_d=\Gamma(\s0d2+2)(4\pi)^{d/2-1}$ to
remove phase space factors. This does not alter the fixed point structure. The
scaling $g\to g/(384\pi^2)$ reproduces the $4d$ classical force law in the
non-relativistic limit \cite{Robinson:2006yd}.  For the anomalous dimension,
we find
\begin{eqnarray}
\label{eta-inf}
\eta(\lambda, g;d)&=&     
\0{(d+2)\, g}{ g- g_{\rm bound}(\lambda)}\,,
\quad\quad
g_{\rm bound}(\lambda;d)
=
\0{\left(1-2\lambda\right)^2}{2(d-2)}\,.
\end{eqnarray}
\begin{table}[t]
\begin{displaymath}
\begin{array}{c|c|c|c}
&
\quad\quad\theta'\quad\quad&
\quad\quad \theta''\quad\quad &
\quad\quad {\rm ref.}\quad\quad 
\\ \hline
\quad a)\quad&
1.1-2.3 & 
2.5-7.0 & 
\cite{Lauscher:2001ya}
\\
b)&
1.4-2.0 & 
2.4-4.3 & 
\cite{Litim:2003vp}
\\
c)&
1.5-1.7 & 
3.0-3.2 & 
\cite{Fischer:2006fz}
\\
\end{array}
\end{displaymath}
\caption{ \label{tEH-4d}The variation of $4d$ scaling exponents
  $\theta_{1,2}=\theta'\pm i\theta''$ in the Einstein-Hilbert theory with the
  gauge fixing parameter $\alpha$ and the cutoff function $R_k$. Results
  indicate the range covered under $a)$ partial variation of both $\alpha$ and
  $R_k$, $b)$ full $\alpha$-variation with optimised $R_k$, and $c)$ full
  $R_k$-variation and optimisation in Feynman gauge ($\alpha=1$).  In all
  cases the fixed point is stable. The variation with $R_k$, amended by
  stablity considerations \cite{Litim:2000ci,Litim:2001up}, is weaker than the
  $\alpha$-variation.}
\end{table}
The anomalous dimension vanishes for vanishing gravitational coupling, and for
$d=\pm2$.  At a non-trivial fixed point the vanishing of $\beta_g$ implies
$\eta_*=2-d$, and reflects the fact that the gravitational coupling is
dimensionless in two dimensions.  At $g=g_{\rm bound}$, the anomalous
dimension $\eta$ diverges. The full flow \eq{ERG} is finite (no poles) and
well-defined for all $k$, as are the full $\beta$-functions derived from
it. Therefore the curve $g=g_{\rm bound}(\lambda)$ limits the domain of
validity of the approximation.

We first consider the case $\lambda=0$ and find two fixed points, the gaussian
one at $g_*=0$ and a non-gaussian one at $g_*=1/(4d)<g_{\rm bound}(0)$, which
are connected under the renormalisation group.  The universal eigenvalue
$\partial \beta_g/\partial g|_*=-\theta$ at the fixed point are $\theta=2-d$
at the gaussian, and
\beq \label{theta0-NG}
\theta=2d\ \0{d-2}{d+2} 
\eeq
at the non-gaussian fixed point.  

Next we allow for a non-vanishing dynamical
cosmological constant term $\Lambda_k\neq 0$ in \eq{EHk}. The coupled system
exhibits the gaussian fixed point $(\lambda_*,g_*)=(0,0)$ with eigenvalues
$-2$ and $d-2$.
\begin{table}[t]
\begin{displaymath}
\begin{array}{c|c|c|c|c|c|c}
&
\quad i \quad
&\theta'
&\quad \theta''\quad 
&\quad \theta_3\quad 
&\quad \theta_4\quad
&\quad {\rm ref.}\quad 
\\ \hline
\quad a)\quad & 2& \quad 2.1 - 3.4 \quad &\quad 3.1 - 4.3\quad   
&\quad 8.4 - 28.8\quad   &-
&
\cite{Lauscher:2002mb}
\\
b)&2 &1.4  &2.8  &25.6  &-
& 
\cite{DL06}
\\
c)&
2 &1.7  &3.1  &3.5  & -
&
\cite{DL06}
\\
d)&
2 &1.4  &2.3  &26.9  & -
& 
\cite{Codello:2007bd}
\\
e)&
3 &2.7  &2.3  &2.1  & -4.2  
&
\cite{Codello:2007bd}
\\
f)&
4 &2.9  &2.5  &1.6  & -3.9  
&
\cite{Codello:2007bd}
\\
g)&
5 &2.5  &2.7  &1.8  & -4.4 
&
\cite{Codello:2007bd}
\\
h)&
6 &2.4  &2.4  &1.5  &  -4.1 
&
\cite{Codello:2007bd}
\\
\end{array}
\end{displaymath}
\caption{\label{tRn-4d}The variation of $4d$ scaling exponents with the order
  $i$ of the expansion, including the invariants $\int\sqrt{\det g_{\mu\nu}}$
  and $\int\sqrt{\det g_{\mu\nu}} R^i$ from $i=2\cdots 6$, using $a)$ Feynman
  gauge under partial variations of $R_k$; $b)$ and $c)$ Feynman gauge and
  optimised $R_k$; $d)-h)$ Landau-deWitt background field gauge with optimised
  $R_k$.}
\end{table}
Non-trivial fixed points of \eq{beta-l-inf} and \eq{beta-g-inf} are found as
follows: For non-vanishing $\lambda\neq 0$, we find a non-trivially vanishing
$\beta_g$ for $g=g_0(\lambda)$, with $g_0(\lambda)=\s01{4d}(1-2\lambda)^2$.
Note that $g_0(\lambda)<g_{\rm bound}(\lambda)$ for all $d>2$.  Evaluating
\eq{beta-l-inf} for $g=g_0(\lambda)$, we find
$\beta_\lambda(\lambda,g_0(\lambda))=
\s014(d-4)(d+1)(1-2\lambda)^2-2d\,\lambda+\s0d2\,.  $ The first term vanishes
in $d=4$ dimensions. Consequently, we find a unique ultraviolet fixed point
$\lambda_*=\s014$ and $g_*=\s01{64}$. In $d> 4$, the vanishing of
$\beta_\lambda$ leads to two branches of real fixed points with either
$\lambda_*>\s012$ or $\s012>\lambda_*>0$. Only the second branch corresponds
to an ultraviolet fixed point which is connected under the renormalisation
group with the correct infrared behaviour. This can be seen as follows: At
$\lambda=\s012$, we find $\eta=d+2>0$. On a non-gaussian fixed point, however,
$\eta<0$. Furthermore, $g$ cannot change sign under the renormalisation group
flow \eq{beta-g-inf}. Consequently, $\eta$ cannot change sign either. Hence,
to connect a fixed point at $\lambda>\s012$ with the gaussian fixed point at
$\lambda=0$, $\eta$ would have to change sign at least twice, which is
impossible. Therefore, we have a unique physically relevant solution given by
\begin{equation}\label{FP-d}
\lambda_*=
\frac{d^2-d-4-\sqrt{2d(d^2-d-4)}}{2(d-4)(d+1)}\,,\quad\quad
g_*=
\0{(\sqrt{d^2-d-4}-\sqrt{2d})^2}{2(d-4)^2(d+1)^2}\,.
\end{equation}
An interesting property of this system is that the scaling exponents
$\theta_1$ and $\theta_2$ -- the eigenvalues of the stability matrix
$\partial\beta_i/\partial g_j$ $(g_1\equiv g,g_2\equiv\lambda)$ at the fixed
point -- are a complex conjugate pair, $\theta_{1,2}=\theta'\pm i\theta''$ with
$\theta'=\s053$ and $\theta''=\s0{\sqrt{167}}{3}$ in four dimensions. The
reason for this is that the stability matrix, albeit real, is not
symmetric. Complex eigenvalues reflect that the interactions at the fixed
point have modified the scaling behaviour of the underlying operators
$\int\sqrt{\det g_{\mu\nu}}R$ and $\int\sqrt{\det g_{\mu\nu}}$. This pattern
changes for lower and higher dimensions, where eigenvalues are real
\cite{Litim:2006dx}.

At this point it is important to check whether the fixed point structure and
the scaling exponents depend on technical parameters such as the gauge fixing
procedure or the momentum cutoff function $R_k$, see Tab.~\ref{tEH-4d} and
~\ref{tRn-4d}. For the Einstein-Hilbert theory in $4d$, results are summarised
in Tab.~\ref{tEH-4d}. The $\alpha$-dependence of the $\beta$-functions is
fairly non-trivial, $e.g.$
\cite{Reuter:1996cp,Lauscher:2001ya,Litim:2003vp}. It is therefore noteworthy
that scaling exponents only depend mildly on variations thereof. Furthermore,
the $R_k$-dependence is smaller than the dependence on gauge fixing
parameters. We conclude that the fixed point is fully stable and
$R_k$-independent for all technical purposes, with the presently largest
uncertainty arising through the gauge fixing sector.  In Tab.~\ref{tRn-4d}, we
discuss the stability of the fixed point under extensions beyond the
Einstein-Hilbert approximation, including higher powers of the Ricci scalar
both in Feynman gauge \cite{Lauscher:2002mb,DL06} and in Landau-DeWitt gauge
\cite{Codello:2007bd}. Once more, the fixed point and the scaling exponents
come out very stable. Furthermore, starting from the operator $\int \sqrt{\det
  g_{\mu\nu}}R^3$ and higher, couplings become irrelevant with negative
scaling exponents \cite{Codello:2007bd,Machado:2007ea}. This is an important
first indication for the set of relevant operators at the UV fixed point being
finite. Finally, we mention that the stability of the fixed point under the
addition of non-interacting matter fields has been confirmed in
\cite{Percacci:2002ie}.

  \begin{table}
\begin{center}
\begin{tabular}{c|ccccccc}
$d$  & 5 & 6 & 7 & 8 & 9 & 10
\\
\hline
$\theta'$ 
&2.69 -- 3.11&4.26 -- 4.78&6.43 -- 6.89&8.19 -- 
                   9.34&10.5 -- 12.1&13.1 -- 15.2
\\
$\theta''$
&4.54 -- 5.16&6.52 -- 7.46&8.43 -- 9.46&10.3 -- 
                   11.4&12.1 -- 13.2&13.9 -- 15.0
\\
$|\theta|$
&5.31 -- 6.06&7.79 -- 8.76&10.4 -- 11.6&13.2 -- 
                   14.7&16.1 -- 17.9&19.1 -- 21.3
\\
\end{tabular}
\end{center}
\caption{The variation of scaling exponent with dimensionality, gauge fixing
  parameters (using either Feynman gauge, or harmonic background field gauge
  with $0\leq \alpha \leq 1$), and the regulator $R_k$; data from
  \cite{Fischer:2006fz,Fischer:2006at}. The $R_k$-variation, covering various
  classes of cutoff functions, is on the level of a few percent and smaller
  than the variation with $\alpha$.}
\label{tEH-d}
    \end{table}

\section{Extra Dimensions}
\label{ED}

It is interesting to discuss fixed points of quantum gravity specifically in
more than four dimensions. The motivation for this is that, first of all, the
critical dimension of gravity -- the dimension where the gravitational
coupling has vanishing canonical mass dimension -- is two. For any dimension
above the critical one, the canonical dimension is negative. Hence, from a
renormalisation group point of view, the four-dimensional theory is not
special. Continuity in the dimension suggests that an ultraviolet fixed point,
if it exists in four dimensions, should persist towards higher dimensions.
More generally, one expects that the local structure of quantum fluctuations,
and hence local renormalisation group properties of a quantum theory of
gravity, are qualitatively similar for all dimensions above the critical one,
modulo topological effects for specific dimensions.  Secondly, the dynamics of
the metric field depends on the dimensionality of space-time. In four
dimensions and above, the metric field is fully dynamical. Hence, once more,
we should expect similarities in the ultraviolet behaviour of gravity in four
and higher dimensions.  Interestingly, this pattern is realised in the results
\cite{Litim:2003vp}, see the analytical fixed point \eq{FP-d}. An extended
systematic search for fixed points in higher-dimensional gravity for general
cutoff $R_k$ has been presented in \cite{Fischer:2006fz,Fischer:2006at}, also
testing the stability of the result against variations of the gauge fixing
parameter (see Tab.~\ref{tEH-d}). The variation with $R_k$,
ammended by stability considerations, is smaller than the variation with
$\alpha$. We conclude from the weak variation that the fixed point indeed
persists in higher dimensions. Further studies including higher derivative
operators confirm this picture \cite{DL06}. This structural stability also
strengthens the results in the four-dimensional case, and supports the view
introduced above. A phenomenological application of these findings in
low-scale quantum gravity is discussed below (see Sec.~\ref{pheno}).

    \begin{figure}
\begin{center}
\unitlength0.001\hsize
\begin{picture}(700,480)
\put(750,20){\Large $\lambda_{4d}$}
\put(-70,410){\Large $g_{4d}$}
\includegraphics[width=.7\hsize]{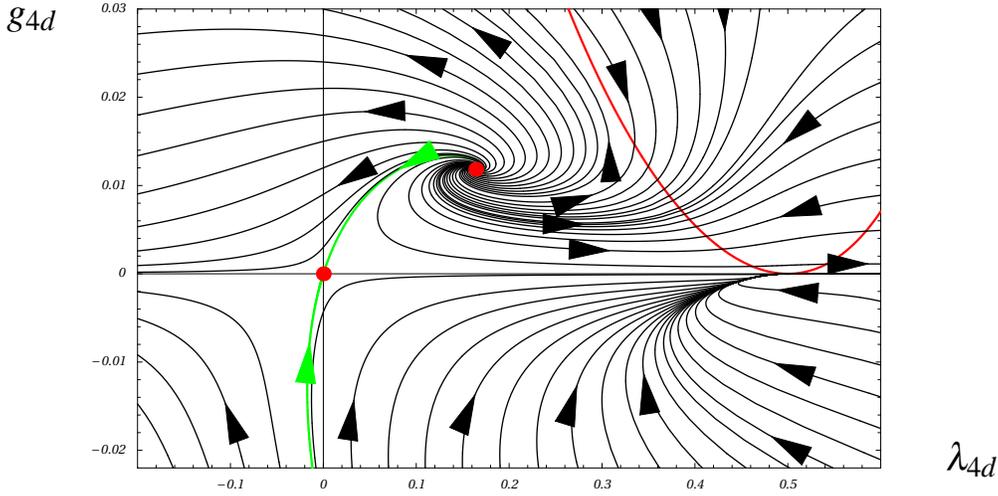}
\end{picture}
\caption{ The phase diagram for the running gravitational coupling
  $g_{4d}$ and the cosmological constant $\lambda_{4d}$ in four dimensions. The
  Gaussian and the ultraviolet fixed point are indicated by dots (red). The
  separatrix connects the two fixed points (full green line). 
The full (red) line indicates the bound $g_{\rm bound}(\lambda)$ where
  $1/\eta=0$. Arrows indicate the direction of the RG flow with decreasing
  $k\to 0$.}
\label{PD4}
\end{center}
\end{figure}

\section{Phase Diagram}
\label{PD}
In this section, we discuss the main characteristics of the phase portrait of
the Einstein-Hilbert theory \cite{Reuter:2001ag,Litim:2003vp} (see
Fig.~\ref{PD4}).  Finiteness of the flow \eq{ERG} implies that the line
$1/\eta=0$ cannot be crossed.  Slowness of the flow implies that the line
$\eta=0$ can neither be crossed (see Sec.~\ref{FP}).  Thus, disconnected
regions of renormalisation group trajectories are characterised by whether $
g$ is larger or smaller $ g_{\rm bound}$ and by the sign of $g$. Since $\eta$
changes sign only across the lines $\eta=0$ or $1/\eta=0$, we conclude that
the graviton anomalous dimension has the same sign along any trajectory. In
the physical domain which includes the ultraviolet and the infrared fixed
point, the gravitational coupling is positive and the anomalous dimension
negative. In turn, the cosmological constant may change sign on trajectories
emmenating from the ultraviolet fixed point.  Some trajectories terminate at
the boundary $g_{\rm bound}(\lambda)$, linked to the present
approximation. The two fixed points are connected by a separatrix.  The
rotation of the separatrix about the ultraviolet fixed point reflects the
complex nature of the eigenvalues. At $k\approx M_{\rm Pl}$, the flow displays
a crossover from ultraviolet dominated running to infrared dominated
running. The non-vanishing cosmological constant modifies the flow mainly in
the crossover region rather than in the ultraviolet.  In the infrared limit,
the separatrix leads to a vanishing cosmological constant $\Lambda_k=\lambda_k
k^2\to 0$ and is interpreted as a phase transition boundary between
cosmologies with positive or negative cosmological constant at large
distances. Trajectories in the vicinity of the separatrix lead to a
positive cosmological constant at large scales and are, therefore, candidate
trajectories for realistic cosmologies \cite{cosmology}.  This picture agrees
very well with numerical results for a sharp cut-off flow
\cite{Reuter:2001ag}, except for the location of the line $1/\eta=0$ which is
non-universal. Similar phase diagrams are found in higher dimensions
\cite{Fischer:2006fz,Fischer:2006at}.

\section{Lattice}
\label{Lattice}
Lattice implementations for gravity in four dimensions have been put forward
based on Regge calculus techniques \cite{Hamber:1999nu,Hamber:2005vc} and
causal dynamical triangulations \cite{Ambjorn:2004qm}.  In the Regge calculus
approach, a critical point which allows for a lattice continuum limit has been
given in \cite{Hamber:1999nu} using the Einstein Hilbert action with fixed
cosmological constant. A scaling exponent has been measured in the
four-dimensional simulation based on varying Newton's coupling to the critical
point, with $\partial_g\beta_g|_*=-\s01\nu$. The result reads $\nu\approx
\s013$, and should be contrasted with the RG result $\nu=1/\theta=\s038$
\cite{Litim:2003vp} as discussed in Sec.~\ref{FP}. In the large-dimensional
limit, geometrical considerations on the lattice lead to the estimate
$\nu=\s01{d-1}$ \cite{Hamber:2005vc}, a behaviour which is in qualitative
agreement with the explicit RG fixed point result $\nu=\s01{2d}$ in the
corresponding limit, see \eq{theta0-NG}.

Within the causal dynamical triangulation approach, global aspects of quantum
space-times have been assessed in \cite{Ambjorn:2004qm}. There, the effective
dimensionality of space-time has been measured as a function of the length
scale by evaluating the return probability of random walks on the triangulated
manifolds. The key result is that the measured effective dimensionality
displays a cross-over from $d\approx 4$ at large scales to $d\approx 2$ at
small scales of the order of the Planck scale. This behaviour compares nicely
with the cross-over of the graviton anomalous dimension $\eta$ under the
renormalisation group (see Sec.~\ref{AS}), and with renormalisation group
studies of the spectral dimension (see
\cite{Niedermaier:2006ns,NiedermaierReuter,Lauscher:2005qz}). These findings
corroborate the claim that asymptotically safe quantum gravity behaves, in an
essential way, two-dimensional at short distances.

\section{Phenomenology}
\label{pheno}
The phenomenology of a gravitational fixed point covers the physics of black
holes \cite{blackholes}, cosmology
\cite{cosmology,Bonanno:2007wg,Bentivegna:2003rr}, modified dispersion
relations \cite{Girelli:2006sc}, and the physics at particle colliders
\cite{Litim:2007iu,Hewett:2007st,Koch:2007yt}.  In this section we concentrate
on the later within low-scale quantum gravity models \cite{add,aadd}.  There,
gravity propagates in $d=4+n$ dimensional bulk whereas matter fields are
confined to a four-dimensional brane. The four-dimensional Planck scale
$M_{\rm Pl}\approx 10^{19}$~GeV is no longer fundamental as soon as the $n$
extra dimensions are compact with radius $\sim L$. Rather, the
$d=4+n$-dimensional Planck mass $M_*$ sets the fundamental scale for gravity,
leading to the relation $M^2_{\rm Pl}\sim M^2_{*} (M_*\,L)^{n}$ for the
four-dimensional Planck scale.  Consequently, $M_*$ can be significantly lower
than $M_{\rm Pl}$ provided $1/L\ll M_*$.  If $M_*$ is of the order of the
electroweak scale, this scenario lifts the hierarchy problem of the standard
model and opens the exciting possibility that particle colliders could
establish experimental evidence for the quantisation of gravity
\cite{grw,tao,virtual_kk}.

\begin{figure}
\begin{picture}(1000,20)
\end{picture}
\unitlength0.001\hsize\includegraphics[width=.4\hsize]{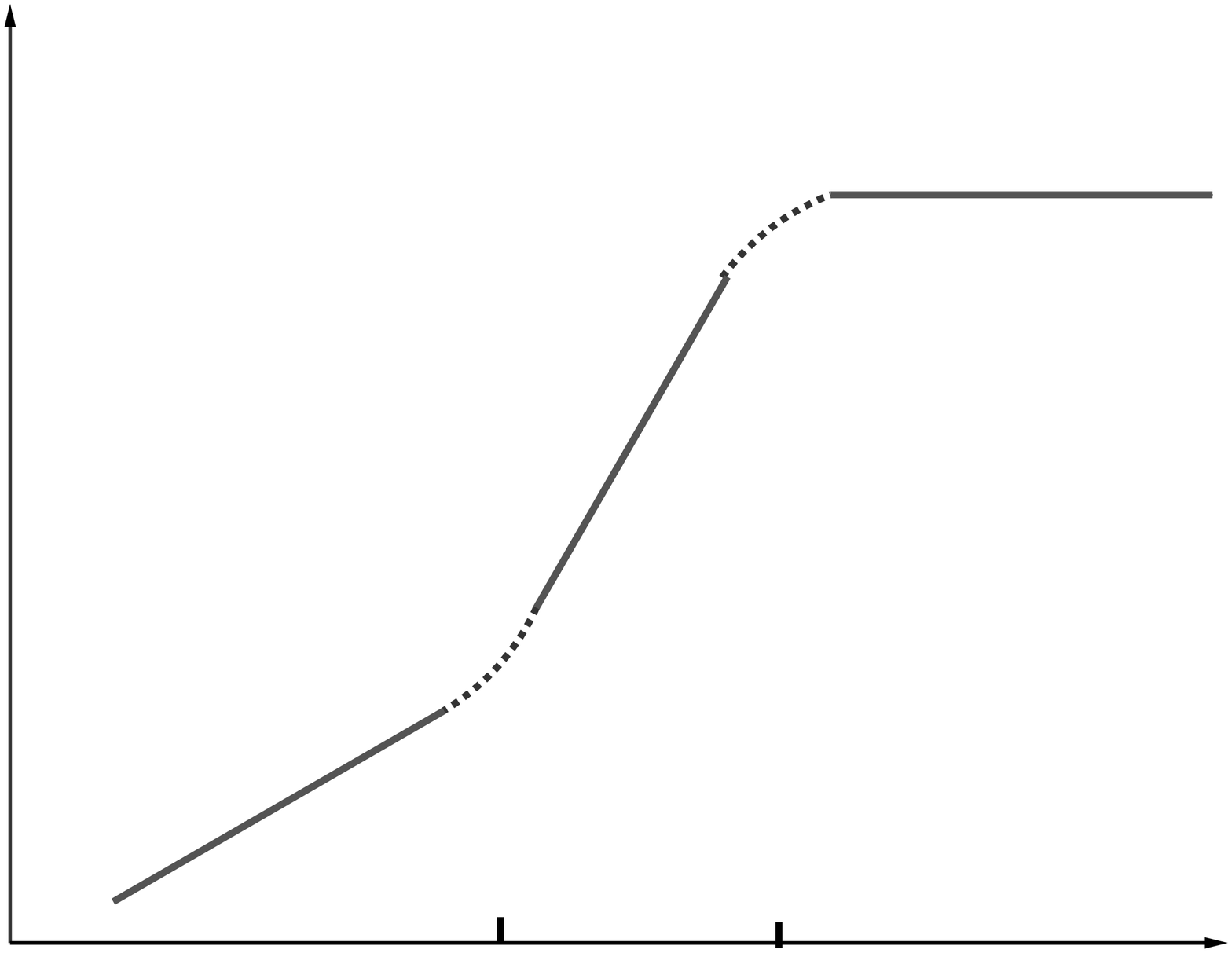}
\vskip-.325\hsize
\hskip.5\hsize
\includegraphics[width=.49\hsize]{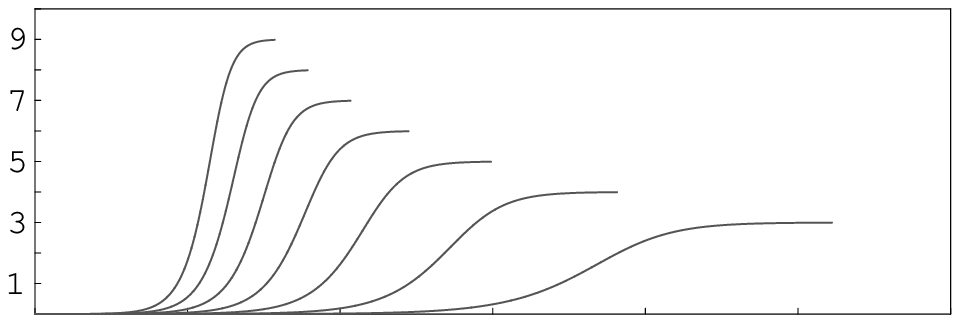}
\vskip-.025\hsize
\hskip.5\hsize
\includegraphics[width=.49\hsize]{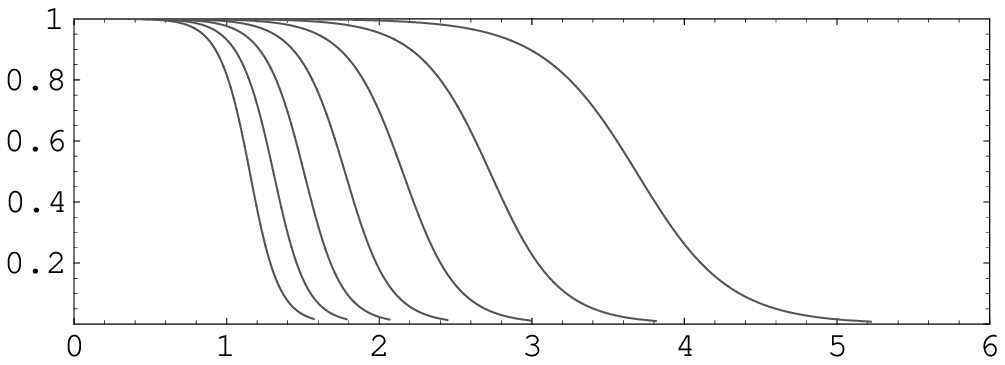}
\begin{picture}(1000,5)
\put(880,320){{$|\eta|$}}
\put(865,265){{$n=1$}}
\put(665,340){{$n=7$}}
\put(880,150){{\large $\displaystyle \frac{G_k}{G_0}$}}
\put(130,390){{a)}\ \ schematically}
\put(660,390){{b)}\ \ numerically}
\put(20,340){$\ln g$}
\put(270,310){UV (fixed point)}
\put(155,200){IR\ \ \ \ \  ($d=4+n$)}
\put(20,80){IR\ \ \ \ \ \ \ \  ($d=4$)}
\put(410,45){$\ln k$}
\put(740,20){$\ln k$}
\put(230,20){$\ln M_*$}
\put(130,20){$\ln 1/L$}
\end{picture}
\caption{ The scale-dependence of the gravitational coupling in a scenario
  with large extra dimension of size $\sim L$ with fundamental Planck scale
  $M_*$ and $M_*L\gg 1$. The fixed point behaviour in the deep ultraviolet
  enforces a softening of gravitational coupling (see text). {a)} In the
  infrared (IR) regime where $|\eta|\ll 1$, the coupling $g=G_k\,k^{d-2}$
  displays a crossover from 4-dimensional to $(4+n)$-dimensional classical
  scaling at $k\approx 1/L$.  The slope ${\rm d}\ln g/{\rm d}\ln k\approx d-2$
  measures the effective number of dimensions. At $k\approx M_*$, a
  classical-to-quantum crossover takes place from $|\eta|\ll1$ to $\eta\approx
  2-d$ (schematically). {b)} Classical-to-quantum crossover at the respective
  Planck scale for $G_k$ and the anomalous dimensions $\eta$ from numerical
  integrations of the flow equation; $d=4+n$ dimensions with $n=1,\cdots,7$
  from right to left.}
\label{RunningG}
\end{figure}

The renormalisation group running of the gravitational coupling in this
scenario has been studied in \cite{Fischer:2006fz,Fischer:2006at,Litim:2007iu}
and is summarised in Fig.~\ref{RunningG}. The main effects due to a fixed
point at high energies set in at momentum scales $k\approx M_*$, where the
gravitational coupling displays a cross-over from perturbative scaling
$G(k)\approx$ const.~to fixed point scaling $G(k)\approx g_*
k^{2-d}$. Therefore we expect that signatures of this cross-over should be
visible in scattering processes at particle colliders as long as these are
sensitive to momentum transfers of the order of $M_*$.

We illustrate this at the example of dilepton production through virtual
gravitons at the Large Hadron Collider (LHC) \cite{Litim:2007iu}.  To lowest
order in canonical dimension, the dilepton production amplitude is generated
through an effective dimension--8 operator in the effective action, involving
four fermions and a graviton~\cite{grw}.  Tree--level graviton exchange is
described by an amplitude ${A} = {S}\cdot {T}$, where ${T} = T_{\mu\nu}
T^{\mu\nu} - \frac{1}{2+n} T_\mu^\mu T_\nu^\nu$ is a function of the
energy-momentum tensor, and
\begin{equation} \label{S}
{S}= \frac{2\pi^{n/2}}{\Gamma(n/2)} \; 
  \frac{1}{M_*^{4}} \;   
\int_0^\infty \frac{d m}{M_*} \; \left(\frac{m}{M_*}\right)^{n-1}\, 
{\cal G}(s,m)
\end{equation}
is a function of the scalar part ${\cal G}(s,m)$ of the graviton
propagator~\cite{grw,gps}.  The integration over the Kaluza-Klein masses $m$,
which we take as continuous, reflects that gravity propagates in the
higher-dimensional bulk. If the graviton anomalous dimension is small, the
propagator is well approximated by ${\cal G}(s,m)=(s+m^2)^{-1}$. This
propagator is used within effective theory settings, and applicable if the
relevant momentum transfer is $\ll M_*$. In this case, \eq{S} is ultraviolet
divergent for $n\ge 2$ due to the Kaluza-Klein modes
\cite{grw}. Regularisation by an UV cutoff leads to a power-law dependence of
the amplitude ${S}\sim M_*^{-4}( {\Lambda}/{M_*} )^{n-2}$ on the cutoff
$\Lambda$.  In a fixed point scenario, the behaviour of $S$ is improved due to
the non-trivial anomalous dimension $\eta$ of the graviton, $e.g.$
\eq{eta-inf}. Evaluating $\eta$ at momentum scale $k^2\approx s+m^2$, we are
lead to the dressed propagator ${\cal
  G}(s,m)\approx\frac{M_*^{n+2}}{(s+m^2)^{n/2+2}}$ in the vicinity of an UV
fixed point. The central observation is that \eq{S} becomes finite even in the
UV limit of the integration. An alternative matching has been adapted in
\cite{Hewett:2007st,Koch:2007yt}, based on the substitution $G(k)\to
G(\sqrt{s})$ in \eq{S}, setting $G=M_*^{2-d}$. In that case, however, \eq{S}
remains UV divergent due to the Kaluza-Klein modes. We conclude that the large
anomalous dimension in asymptotically safe gravity provides for a finite
dilepton production rate.

\begin{figure*}[t]
\begin{picture}(1000,112)
\unitlength0.001\hsize
\put(80,250){{a)}\ \ effective theory}
\put(550,250){{b)}\ \ renormalisation group}
\includegraphics[width=.33\textwidth]{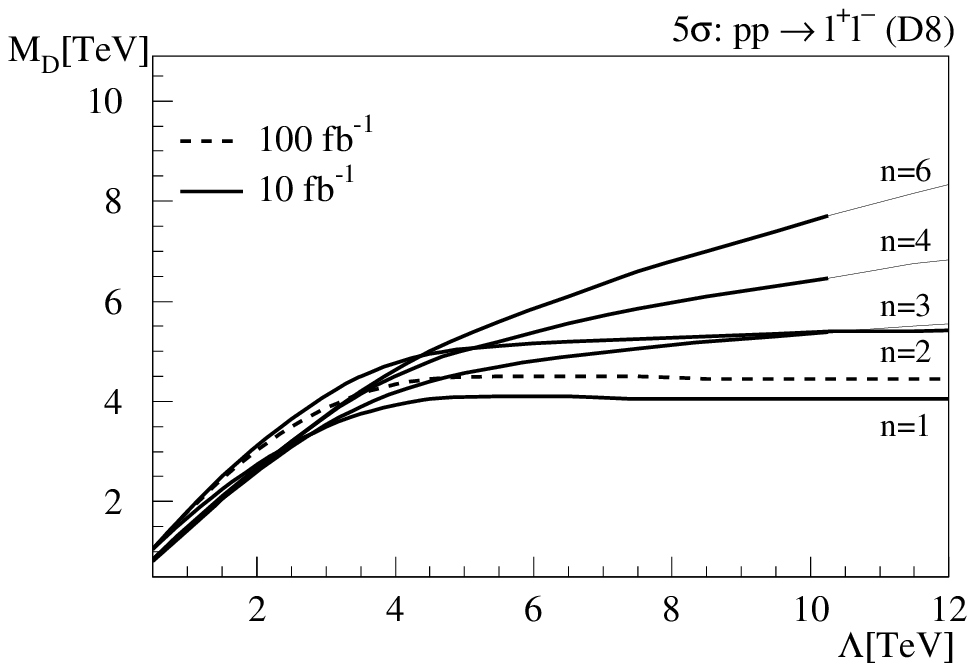}
\includegraphics[width=.66\textwidth]{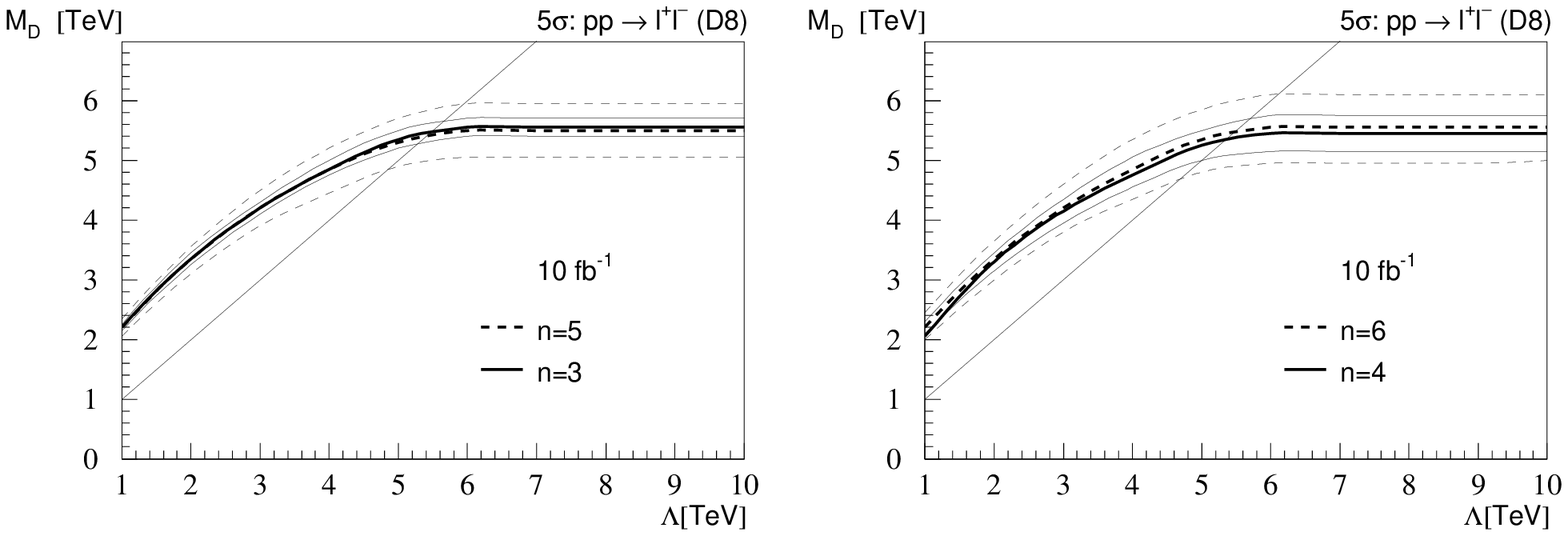}
\end{picture}
\vskip-.3cm
\caption{The $5\sigma$ discovery contours in $M_D$ at the LHC ($d=4+n)$, as a
  function of a cutoff $\Lambda$ on $E_{\rm parton}$ for an assumed integrated
  luminosity of $10{\rm fb}^{-1}$ ($100{\rm fb}^{-1}$). a) Effective theory:
  the sensitivity to the cutoff $\Lambda$ is
  reflected in the $M_D$ contour; plot from \cite{gps}. b) Renormalisation
  group: the limit $\Lambda\to\infty$ can be performed, and the leveling-off
  at $M_D\approx \Lambda$ reflects the gravitational fixed
  point, thin lines show a $\pm$10\% variation in the transition scale; plot
  from \cite{Litim:2007iu}.  }
\label{fig:discovery}
\vskip-.3cm
\end{figure*}

In Fig.~\ref{fig:discovery} we show the discovery potential in the fundamental
Planck scale at the LHC, and compare effective theory studies \cite{gps} with
a gravitational fixed point \cite{Litim:2007iu}. In either case the minimal
signal cross sections have been computed for which a $5\sigma$ excess can be
observed, taking into account the leading standard model backgrounds and
assuming statistical errors. This translates into a maximum reach $M_D$ for
the fundamental Planck scale $M_*$.  To estimate uncertainties in the RG
set-up, we allow for a 10\% variation in the scale where the transition
towards fixed point scaling sets in.  Consistency is checked by introducing an
artificial cutoff $\Lambda$ on the partonic energy \cite{gps}, setting the
partonic signal cross section to zero for $E_{\rm parton}>\Lambda$.  It is
nicely seen that $M_D$ becomes independent of $\Lambda$ for $\Lambda\to\infty$
when fixed point scaling is taken into account.

\section{Conclusions}
\label{conclusions}
The asymptotic safety scenario offers a genuine path towards quantum gravity
in which the metric field remains the fundamental carrier of the physics even
in the quantum regime. We have reviewed the ideas behind this set-up in the
light of recent advances based on renormalisation group and lattice studies.
The stability of renormalisation group fixed points and scaling exponents
detected in four- and higher-dimensional gravity is remarkable, strongly
supporting this scenario.  Furthermore, underlying expansions show good
numerical convergence, and uncertainties which arise through approximations
are moderate.  If the fundamental Planck scale is as low as the electroweak
scale, signs for the quantisation of gravity and asymptotic safety could even
be observed in collider experiments. It is intriguing that key aspects of
asymptotic safety are equally seen in lattice studies. It will be interesting
to evaluate these links more deeply in the future.  Finally, asymptotically
safe gravity is a natural set-up which leads to classical general relativity
as a ``low energy phenomenon'' of a fundamental quantum field theory in the
metric field.

\section*{Acknowledgements}
I thank Peter Fischer and Tilman Plehn for collaboration on the topics
discussed here, and the organisers for their invitation to a very stimulating
workshop.



\begin{thebibliography}{99}

\bibitem{Weinberg} S.~Weinberg, in {\it General Relativity: An
Einstein centenary survey}, Eds.~S.W.~Hawking and W.~Israel, Cambridge
University Press (1979), p.~790.

\bibitem{Litim:2006dx}
D.~F.~Litim,
  {\it On fixed points of quantum gravity},
  AIP Conf.\ Proc.\  {\bf 841} (2006) 322 [hep-th/0606044].

\bibitem{Niedermaier:2006ns}
  M.~Niedermaier,
  {\it The asymptotic safety scenario in quantum gravity: An introduction,}
  Class.\ Quant.\ Grav.\  {\bf 24} (2007) R171
  [gr-qc/0610018].

\bibitem{NiedermaierReuter} M.~Niedermaier and M.~Reuter, {\it The Asymptotic
  Safety Scenario in Quantum Gravity}, Living Rev.~Relativity {\bf 9} (2006) 5.

\bibitem{Percacci:2007sz}
  R.~Percacci,
  {\it Asymptotic Safety,}
in 'Approaches to Quantum Gravity: Towards a New Understanding of Space, Time and Matter' ed. D. Oriti, Cambridge University Press,  0709.3851 [hep-th].




\bibitem{Reuter:1996cp}
M.~Reuter,
{\it Nonperturbative Evolution Equation for Quantum Gravity},
Phys.\ Rev.\ D {\bf 57} (1998) 971 [hep-th/9605030].
%

\bibitem{Souma:1999at}
W.~Souma,
{\it Non-trivial ultraviolet fixed point in quantum gravity},
Prog.\ Theor.\ Phys.\  {\bf 102} (1999) 181 [hep-th/9907027];
  {\it Gauge and cutoff function dependence of the ultraviolet fixed point in
  quantum gravity},
  gr-qc/0006008.

\bibitem{Lauscher:2001ya}
O.~Lauscher and M.~Reuter,
{\it Ultraviolet fixed point and generalized flow equation of quantum  gravity},
Phys.\ Rev.\ D {\bf 65} (2002) 025013 [hep-th/0108040].

\bibitem{Lauscher:2002mb}
O.~Lauscher and M.~Reuter,
{\it Is quantum Einstein gravity nonperturbatively renormalizable?},
Class.\ Quant.\ Grav.\  {\bf 19} (2002) 483 [hep-th/0110021];
%
{\it Flow equation of quantum Einstein gravity in a higher-derivative truncation},
Phys.\ Rev.\ D {\bf 66} (2002) 025026 [hep-th/0205062].

\bibitem{Reuter:2001ag}
M.~Reuter and F.~Saueressig,
{\it Renormalization group flow of quantum gravity in the Einstein-Hilbert truncation},
Phys.\ Rev.\ D {\bf 65} (2002) 065016 [hep-th/0110054].

\bibitem{Percacci:2002ie}
R.~Percacci and D.~Perini,
{\it Constraints on matter from asymptotic safety},
Phys.\ Rev.\ D {\bf 67} (2003) 081503 [hep-th/0207033];
%
{\it Asymptotic safety of gravity coupled to matter},
Phys.\ Rev.\ D 
{\bf 68} (2003) 044018 [hep-th/0304222].




\bibitem{Litim:2003vp}
  D.~F.~Litim,
  {\it Fixed points of quantum gravity},
  Phys.\ Rev.\ Lett.\  {\bf 92} (2004) 201301
  [hep-th/0312114].


\bibitem{Bonanno:2004sy}
  A.~Bonanno and M.~Reuter,
  {\it Proper time flow equation for gravity},
  JHEP {\bf 0502} (2005) 035
  [hep-th/0410191].


\bibitem{Fischer:2006fz}
  P.~Fischer and D.~F.~Litim,
  {\it Fixed points of quantum gravity in extra dimensions},
  Phys.\ Lett.\  B {\bf 638} (2006) 497
  [hep-th/0602203].

\bibitem{Fischer:2006at}
 P.~Fischer and D.~F.~Litim,
  {\it Fixed points of quantum gravity in higher dimensions},
  AIP Conf.\ Proc.\  {\bf 861} (2006) 336
  [hep-th/0606135].


\bibitem{DL06} D. Litim, unpublished notes.

\bibitem{Codello:2007bd}
  A.~Codello, R.~Percacci and C.~Rahmede,
  {\it Ultraviolet properties of f(R)-gravity},
  Int.\ J.\ Mod.\ Phys.\  A {\bf 23} (2008) 143
  [0705.1769 [hep-th]];
  {\it Investigating the Ultraviolet Properties of Gravity with a Wilsonian
  Renormalization Group Equation},
  0805.2909 [hep-th].

\bibitem{Machado:2007ea}
  P.~F.~Machado and F.~Saueressig,
  {\it On the renormalization group flow of f(R)-gravity},
  Phys.\ Rev.\  D {\bf 77} (2008) 124045
  [0712.0445 [hep-th]].

\bibitem{Forgacs:2002hz}
P.~Forgacs and M.~Niedermaier,
{\it A fixed point for truncated quantum Einstein gravity},
hep-th/0207028.

\bibitem{Niedermaier:2002eq}
M.~Niedermaier,
{\it On the renormalization of truncated quantum Einstein gravity},
JHEP {\bf 0212} (2002) 066 [hep-th/0207143],
{\it Dimensionally reduced gravity theories are asymptotically safe},
Nucl.\ Phys.\ B {\bf 673} (2003) 131 [hep-th/0304117].



\bibitem{Gastmans:1977ad}
  R.~Gastmans, R.~Kallosh and C.~Truffin,
  {\it Quantum Gravity Near Two-Dimensions},
  Nucl.\ Phys.\ B {\bf 133} (1978) 417

\bibitem{Christensen:1978sc}
  S.~M.~Christensen and M.~J.~Duff,
  {\it Quantum Gravity In Two + Epsilon Dimensions},
  Phys.\ Lett.\ B {\bf 79} (1978) 213.


\bibitem{Kawai:1992fz}
  H.~Kawai, Y.~Kitazawa and M.~Ninomiya,
  {\it Scaling exponents in quantum gravity near two-dimensions},
  Nucl.\ Phys.\ B {\bf 393} (1993) 280
  [hep-th/9206081].

\bibitem{Aida:1996zn}
  T.~Aida and Y.~Kitazawa,
  {\it Two-loop prediction for scaling exponents in (2+epsilon)-dimensional
  quantum gravity},
  Nucl.\ Phys.\ B {\bf 491} (1997) 427
  [hep-th/9609077].

\bibitem{Codello:2006in}
  A.~Codello and R.~Percacci,
  {\it Fixed Points of Higher Derivative Gravity},
  Phys.\ Rev.\ Lett.\  {\bf 97} (2006) 221301
  [hep-th/0607128].

\bibitem{Percacci:2005wu}
  R.~Percacci,
  {\it Further evidence for a gravitational fixed point},
  Phys.\ Rev.\  D {\bf 73} (2006) 041501
  [hep-th/0511177].


\bibitem{Hamber:1999nu}
H.~W.~Hamber,
{\it On the gravitational scaling dimensions},
Phys.\ Rev.\ D {\bf 61} (2000) 124008
[hep-th/9912246];
{\it Phases of 4-d simplicial quantum gravity},
Phys.\ Rev.\ D {\bf 45} (1992) 507;
  H.~W.~Hamber and R.~M.~Williams,
  {\it Non-perturbative gravity and the spin of the lattice graviton},
  Phys.\ Rev.\ D {\bf 70} (2004) 124007
  [hep-th/0407039].

\bibitem{Hamber:2005vc}
  H.~W.~Hamber and R.~M.~Williams,
  {\it Quantum gravity in large dimensions},
  Phys.\ Rev.\  D {\bf 73} (2006) 044031
  [hep-th/0512003].



\bibitem{Ambjorn:2004qm}
J.~Ambjorn, J.~Jurkiewicz and R.~Loll,
{\it Emergence of a 4D world from causal quantum gravity},
Phys.\ Rev.\ Lett.\  {\bf 93}, 131301 (2004)
[hep-th/0404156];
  {\it Spectral dimension of the universe},
  Phys.\ Rev.\ Lett.\  {\bf 95} (2005) 171301
  [hep-th/0505113].

\bibitem{Donoghue:1993eb}
  J.~F.~Donoghue,
  {\it Leading quantum correction to the Newtonian potential},
  Phys.\ Rev.\ Lett.\  {\bf 72} (1994) 2996
  [gr-qc/9310024].

\bibitem{Burgess:2003jk} C.~P.~Burgess, {\it Quantum gravity in everyday life:
  General relativity as an effective field theory}, Living Rev.\ Rel.\ {\bf 7}
  (2004) 5 [gr-qc/0311082].


\bibitem{Bergerhoff:1995zq}
  B.~Bergerhoff, F.~Freire, D.~F.~Litim, S.~Lola and C.~Wetterich,
  {\it Phase diagram of superconductors},
  Phys.\ Rev.\ B {\bf 53} (1996) 5734
 [hep-ph/9503334];\\
  B.~Bergerhoff, D.~F.~Litim, S.~Lola and C.~Wetterich,
  {\it Phase transition of N component superconductors},
  Int.\ J.\ Mod.\ Phys.\ A {\bf 11} (1996) 4273
  [cond-mat/9502039].
%
\bibitem{Herbut:1996ut}
  I.~F.~Herbut and Z.~Tesanovic, {\it Critical fluctuations in superconductors
    and the magnetic field penetration depth}, Phys.\ Rev.\ Lett.\ {\bf 76}
  (1996) 4588 [cond-mat/9605185].

\bibitem{Kazakov:2002jd}
  D.~I.~Kazakov,
  {\it Ultraviolet fixed points in gauge and SUSY field theories in extra
  dimensions}
  JHEP {\bf 0303} (2003) 020
  [hep-th/0209100].
%
\bibitem{Gies:2003ic}
  H.~Gies,
  {\it Renormalizability of gauge theories in extra dimensions},
  Phys.\ Rev.\ D {\bf 68} (2003) 085015
  [hep-th/0305208].
\bibitem{Morris:2004mg}
  T.~R.~Morris,
  {\it Renormalizable extra-dimensional models},
  JHEP {\bf 0501} (2005) 002
  [hep-ph/0410142].



\bibitem{Litim:1998yn}
  D.~F.~Litim,
  {\it Wilsonian flow equation and thermal field theory},
  hep-ph/9811272.

\bibitem{Litim:1998nf} D.~F.~Litim and J.~M.~Pawlowski, {\it On gauge
  invariant Wilsonian flows}, in {\it The Exact Renormalization Group},
  Eds.~Krasnitz et al, World Sci (1999) 168 [hep-th/9901063].


\bibitem{Bagnuls:2000ae}
  C.~Bagnuls and C.~Bervillier,
  {\it Exact renormalization group equations: An introductory review},
  Phys.\ Rept.\  {\bf 348} (2001) 91
  [hep-th/0002034].
\bibitem{BTW}
J.~Berges, N.~Tetradis and C.~Wetterich,
{\it Non-perturbative renormalisation flow in quantum field theory 
and  statistical physics},
Phys.\ Rept.\  {\bf 363} (2002) 223 [hep-ph/0005122].
\bibitem{JP}
J.~Polonyi,
{\it Lectures on the functional renormalization group method},
Centr. Eur. Sci. J. Phys.~1 (2002) 1 [hep-th/0110026].

%
\bibitem{Pawlowski:2005xe}
  J.~M.~Pawlowski,
  {\it Aspects of the functional renormalisation group},
  Annals Phys.\  {\bf 322} (2007) 2831
  [hep-th/0512261].

\bibitem{Gies:2006wv}
  H.~Gies,
  {\it Introduction to the functional RG and applications to gauge theories},
  hep-ph/0611146.


\bibitem{Litim:2000ci}
D.~F.~Litim,
{\it Optimisation of the exact renormalisation group},
Phys.\ Lett.\  {\bf B486} (2000) 92 [hep-th/0005245];
{\it Mind the gap},
Int.\ J.\ Mod.\ Phys.\ A {\bf 16} (2001) 2081
[hep-th/0104221];
{\it Convergence and stability of the renormalisation group},
Acta Phys.~Slov.{\bf 52} (2002) 635
[hep-th/0208117].


\bibitem{Wetterich:1992yh}
  C.~Wetterich,
  {\it Exact evolution equation for the effective potential},
  Phys.\ Lett.\  B {\bf 301} (1993)  90.


\bibitem{Litim:2001up}
D.~F.~Litim,
{\it Optimised renormalisation group flows},
Phys.\ Rev.\ D {\bf 64} (2001) 105007 [hep-th/0103195].


\bibitem{Litim:2005us}
  D.~F.~Litim,
  {\it Universality and the renormalisation group},
  JHEP {\bf 0507} (2005) 005
  [hep-th/0503096].



\bibitem{Litim:2001ky}
  D.~F.~Litim and J.~M.~Pawlowski,
  {\it Perturbation theory and renormalisation group equations},
  Phys.\ Rev.\  D {\bf 65} (2002) 081701
  [hep-th/0111191].

\bibitem{Litim:2002xm}
  D.~F.~Litim and J.~M.~Pawlowski,
  {\it Completeness and consistency of renormalisation group flows},
  Phys.\ Rev.\  D {\bf 66} (2002) 025030
  [hep-th/0202188].


\bibitem{Litim:2002cf}
D.~F.~Litim,
{\it Critical exponents from optimised renormalisation group flows},
Nucl.\ Phys.\ B {\bf 631} (2002) 128
[hep-th/0203006];
{\it Derivative expansion and renormalisation group flows},
JHEP {\bf 0111} (2001) 059
[hep-th/0111159].


\bibitem{Pawlowski:2003hq}
  J.~M.~Pawlowski, D.~F.~Litim, S.~Nedelko and L.~von Smekal,
  {\it Infrared behaviour and fixed points in Landau gauge QCD},
  Phys.\ Rev.\ Lett.\  {\bf 93} (2004) 152002
  [hep-th/0312324];



\bibitem{Branchina:2003ek}
  V.~Branchina, K.~A.~Meissner and G.~Veneziano,
  {\it The price of an exact, gauge-invariant RG-flow equation},
  Phys.\ Lett.\  B {\bf 574} (2003) 319
  [hep-th/0309234];\\
  J.~M.~Pawlowski,
  {\it Geometrical effective action and Wilsonian flows},
  hep-th/0310018.

\bibitem{Symanzik:1970rt}
  K.~Symanzik,
  {\it Small Distance Behavior In Field Theory And Power Counting},
  Commun.\ Math.\ Phys.\  {\bf 18} (1970) 227.


\bibitem{Reuter:1993kw}
  M.~Reuter and C.~Wetterich,
  {\it Effective average action for gauge theories and exact evolution
  equations},
  Nucl.\ Phys.\  B {\bf 417} (1994) 181;
{\it Gluon condensation in nonperturbative flow equations},
Phys.\ Rev.\ D {\bf 56} (1997) 7893 [hep-th/9708051];
  H.~Gies,
  {\it Running coupling in Yang-Mills theory: A flow equation study},
  Phys.\ Rev.\  D {\bf 66} (2002) 025006
  [hep-th/0202207].

\bibitem{Litim:1998qi}
  D.~F.~Litim and J.~M.~Pawlowski,
  {\it Flow equations for Yang-Mills theories in general axial gauges},
  Phys.\ Lett.\ B {\bf 435} (1998) 181 [hep-th/9802064];
  {\it Renormalisation group flows for gauge theories in axial gauges},
  JHEP {\bf 0209} (2002) 049
  [hep-th/0203005].

\bibitem{Freire:2000bq}
  F.~Freire, D.~F.~Litim and J.~M.~Pawlowski,
  {\it Gauge invariance and background field formalism in the exact
  renormalisation group},
  Phys.\ Lett.\  B {\bf 495}, 256 (2000) [hep-th/0009110].

\bibitem{Litim:2002hj}
  D.~F.~Litim and J.~M.~Pawlowski,
  {\it Wilsonian flows and background fields},
  Phys.\ Lett.\  B {\bf 546} (2002) 279
  [hep-th/0208216].

\bibitem{Pawlowski:2001df}
  J.~M.~Pawlowski,
  {\it On Wilsonian flows in gauge theories},
  Int.\ J.\ Mod.\ Phys.\  A {\bf 16} (2001) 2105.

\bibitem{Robinson:2006yd}
  S.~P.~Robinson,
  {\it Normalization conventions for Newton's constant and the Planck scale in
  arbitrary spacetime dimension},
  gr-qc/0609060.

\bibitem{Lauscher:2005qz}
  O.~Lauscher and M.~Reuter,
  {\it Fractal spacetime structure in asymptotically safe gravity},
  JHEP {\bf 0510} (2005) 050
  [hep-th/0508202].

\bibitem{blackholes}
  A.~Bonanno and M.~Reuter,
  {\it Quantum gravity effects near the null black hole singularity},
  Phys.\ Rev.\  D {\bf 60} (1999) 084011
  [gr-qc/9811026];
  {\it Renormalization group improved black hole spacetimes},
  Phys.\ Rev.\  D {\bf 62} (2000) 043008
  [hep-th/0002196];
  {\it Spacetime structure of an evaporating black hole in quantum gravity},
  Phys.\ Rev.\  D {\bf 73} (2006) 083005
  [hep-th/0602159].


\bibitem{cosmology}
M.~Reuter and F.~Saueressig, {\it From big bang to asymptotic de Sitter:
  Complete cosmologies in a quantum gravity framework}, JCAP {\bf 0509} (2005)
012 [hep-th/0507167].

\bibitem{Bonanno:2007wg}
  A.~Bonanno and M.~Reuter,
  {\it Entropy signature of the running cosmological constant},
  JCAP {\bf 0708} (2007) 024
  [0706.0174 [hep-th]].

\bibitem{Bentivegna:2003rr} E.~Bentivegna, A.~Bonanno and M.~Reuter, {\it
  Confronting the IR Fixed Point Cosmology with High Redshift Supernova Data},
  JCAP {\bf 0401} (2004) 001 [astro-ph/0303150].


\bibitem{Girelli:2006sc}
  F.~Girelli, S.~Liberati, R.~Percacci and C.~Rahmede,
  {\it Modified dispersion relations from the renormalization group of
    gravity,}, Class.\ Quant.\ Grav.\  {\bf 24} (2007) 3995
  [gr-qc/0607030].


\bibitem{Litim:2007iu}
  D.~F.~Litim and T.~Plehn,
  {\it Signatures of gravitational fixed points at the LHC},
  Phys.\ Rev.\ Lett.\  {\bf 100} (2008) 131301
  [0707.3983 [hep-ph]];
  {\it Virtual Gravitons at the LHC},
  0710.3096 [hep-ph].


\bibitem{Hewett:2007st}
  J.~Hewett and T.~Rizzo,
  {\it Collider Signals of Gravitational Fixed Points},
  JHEP {\bf 0712} (2007) 009
  [0707.3182 [hep-ph]].


\bibitem{Koch:2007yt} B.~Koch, {\it Renormalization group and black hole
  production in large extra dimensions}, Phys.\ Lett.\ B {\bf 663/4} (2008)
  334 [0707.4644 [hep-ph]].





\bibitem{add}
 N.~Arkani-Hamed, S.~Dimopoulos and G.~R.~Dvali,
  {\it The hierarchy problem and new dimensions at a millimeter},
  Phys.\ Lett.\ B {\bf 429} (1998) 263 [hep-ph/9803315].
\bibitem{aadd}
 I.~Antoniadis, N.~Arkani-Hamed, S.~Dimopoulos and G.~R.~Dvali,
  {\it New dimensions at a millimeter to a Fermi and superstrings at a TeV},
  Phys.\ Lett.\ B {\bf 436} (1998) 257 [hep-ph/9804398].

\bibitem{grw}
 G.~F.~Giudice, R.~Rattazzi and J.~D.~Wells,
  {\it Quantum gravity and extra dimensions at high-energy colliders},
  Nucl.\ Phys.\ B {\bf 544} (1999) 3
  [hep-ph/9811291].
\bibitem{tao}
 T.~Han, J.~D.~Lykken and R.~J.~Zhang,
  {\it On Kaluza-Klein states from large extra dimensions},
  Phys.\ Rev.\ D {\bf 59} (1999) 105006,
  [hep-ph/9811350].
\bibitem{virtual_kk}
 J.~L.~Hewett,
  {\it Indirect collider signals for extra dimensions},
  Phys.\ Rev.\ Lett.\  {\bf 82} (1999) 4765,
  [hep-ph/9811356].


\bibitem{gps}
 G.~F.~Giudice, T.~Plehn and A.~Strumia,
  {\it Graviton collider effects in one and more large extra dimensions},
  Nucl.\ Phys.\ B {\bf 706} (2005) 455
  [hep-ph/0408320].

\end{thebibliography}
\end{document}